\documentclass[review,authoryear]{elsarticle}
\graphicspath{ {../Figures/} } 
\usepackage[labelfont=bf]{caption}
\usepackage{lineno,hyperref,natbib,multirow}
\journal{arXiv}

\begin{document}

\begin{frontmatter}

\title{Fast and Accurate Semi-Automatic Segmentation Tool for Brain Tumor MRIs}

\author[systems]{Andrew X. Chen}
\author[systems,dbmi]{Ra\'{u}l Rabad\'{a}n\corref{mycorrespondingauthor}}
\cortext[mycorrespondingauthor]{Corresponding author}
\ead{rr2579@cumc.columbia.edu}

\address[systems]{Department of Systems Biology, Columbia University Medical Center, 1130 St. Nicholas Avenue, New York, NY 10032, USA}
\address[dbmi]{Department of Biomedical Informatics, Columbia University Medical Center, 622 W. 168th Street, New York, NY 10032, USA}

\begin{abstract} 
Segmentation, the process of delineating tumor apart from healthy tissue, is a vital part of both the clinical assessment and the quantitative analysis of brain cancers. Here, we provide an open-source algorithm (MITKats), built on the Medical Imaging Interaction Toolkit, to provide user-friendly and expedient tools for semi-automatic segmentation. To evaluate its performance against competing algorithms, we applied MITKats to 38 high-grade glioma cases from publicly available benchmarks. The similarity of the segmentations to expert-delineated ground truths approached the discrepancies among different manual raters, the theoretically maximal precision. The average time spent on each segmentation was 5 minutes, making MITKats between 4 and 11 times faster than competing semi-automatic algorithms, while retaining similar accuracy.\\
\end{abstract}


\end{frontmatter}
\newpage
%\linenumbers

\section{Introduction}
Glioblastoma (GBM) is the most common primary brain malignancy, and the one with the most dismal prognosis \citep{DeAngelis2001}. Imaging, particularly magnetic resonance imaging (MRI), is the standard for diagnosing and assessing the disease \citep{Mabray2015}. The delineation of tumor volumes within images, or segmentation, is important for guiding therapy \citep{Dupont2016, Bauer2013a}, determining prognosis \citep{Kickingereder2016, Cui2015}, and assessing response\citep{Chow2014, Clarke1998}. Related quantitative spatial analyses have demonstrated utility in predicting molecular subtypes \citep{Yang2016} and survival \citep{Czarnek2017, Mazurowski2014, Zhang2014a} of glioblastoma patients. Given the availability of public imaging datasets such as The Cancer Imaging Archive \citep{Clark2013}, there is a clear need for the development of efficient and accurate segmentation utilities, which will allow for the systematic quantification of images in association with clinical and molecular characteristics.

\subsection{Related Work}
Manual segmentation is typically performed with 2D tools to delineate edges on each image slice. While manual segmentation is the gold standard \citep{Porz2014}, it is too time-consuming for large analyses, sometimes taking upwards of an hour per image series \citep{Kaus2001}. Because of this, many fully automatic and semi-automatic algorithms have been created to expedite the segmentation process. Reviews of these methods can be found in \citet{Bauer2013} and \citet{Wang2014}. Benchmarks of fully automatic algorithms have demonstrated encouraging accuracies \citep{Menze2015}, but acceptance of these methods in the clinic is limited due to concerns about errors and transparency \citep{Gordillo2013}. Semi-automatic algorithms draw a balance by unifying the power of computer processing with the intuition of the human operator. However, existing semi-automatic programs still need improvement with regards to operator time \citep{Fyllingen2016} and user-friendliness \citep{Ramkumar2016}.

\subsection{Contribution}
To address these drawbacks, this work provides and validates an accessible semi-automatic protocol for the fast segmentation of glioblastomas. Specific goals included the creation of an intuitive computer-assisted segmentation utility, addition of 3D editing tools for manual correction, and integration within a user-friendly environment. These aims were implemented in the Medical Imaging Interaction Toolkit (MITK) as an extension of its existing Segmentation plugin \citep{Wolf2005, Maleike2009}. This modified software, MITK with augmented tools for segmentation (MITKats), is also a free and open-source program. By addressing the aforementioned goals with MITKats, we expedite semi-automatic segmentation by introducing validated, easy-to-use 3D tools.

\section{Methods}

\subsection{Data sources} 
A total of 38 3D MRIs of brain tumors were obtained from two publicly available datasets. 20 cases of high-grade gliomas (including glioblastomas and anaplastic astrocytomas) were obtained from the Multimodal Brain Tumor Segmentation Challenge (BRATS) 2012 \citep{Menze2015}, a notable accuracy benchmark. Four modalities were available for the BRATS dataset: T1, T1 with contrast (T1c), T2, and T2 FLAIR, though MITKats only used T1c and FLAIR images. The ground truths were derived from all four modalities and designated into 3 regions: 

\begin{itemize}
	\item Active: The contrast-enhancing parts of the tumor, seen on T1c
	\item Core: The Active component plus non-enhancing features seen on T1, including necrosis
	\item Whole: The hyper-intense region on T2 and FLAIR, corresponding to edema
\end{itemize}

18 cases of glioblastoma were obtained from a previously published study performed at St. Olav's University Hospital \citep{Fyllingen2016}. Only T1c MRIs were used to create segmentations, which were generated in BrainVoyager \citep{Goebel2006}, 3D Slicer \citep{Egger2013}, and ITK-SNAP \citep{Yushkevich2006}. A single tumor region was labeled, comprising enhancing and necrotic regions. Given that the authors did not observe non-enhancing tumor regions, their definition of tumor was closest to that of the Core component from BRATS. While no ground truths were designated in this dataset, each segmentation was timed, serving as a speed benchmark.
 
Both datasets had originally been pre-processed, resulting in interpolation of image resolutions to a 1mm isotropic voxel size. Images from BRATS had been skull-stripped, while those from St. Olav's had not.

\subsection{Algorithmic Development} 
Two modifications were made to the MITK framework in order to expedite the segmentation process in MITKats. First, the Threshold Components tool was added, which expanded connected threshold segmentation to accept multiple seed points (and thus separated regions). It also allowed for the manipulation of seed points independently of the thresholds, as well as streamlining out unnecessary user input. The second modification was adding in segmentation capability to the Clipping Plane View, which was originally intended for volume measurements only. Therefore, a segmentation could be graphically adjusted in three dimensions through extraction of a clipped piece. 

MITKats can be found here: https://github.com/RabadanLab/MITKats, and is in the process of being merged onto the main branch of MITK.

\subsection{Segmentation Protocol} 
Segmentations were performed in MITKats by A.X.C., a medical student who had used similar software to segment 93 glioblastoma cases as part of a previous project \citep{Crawford}. Segmentation time, as measured by stopwatch, was started after loading the original image(s) and stopped after opening the save segmentation dialog, thus ignoring the time for file operations. 

\begin{figure}[htp!]
	\includegraphics[width=\textwidth]{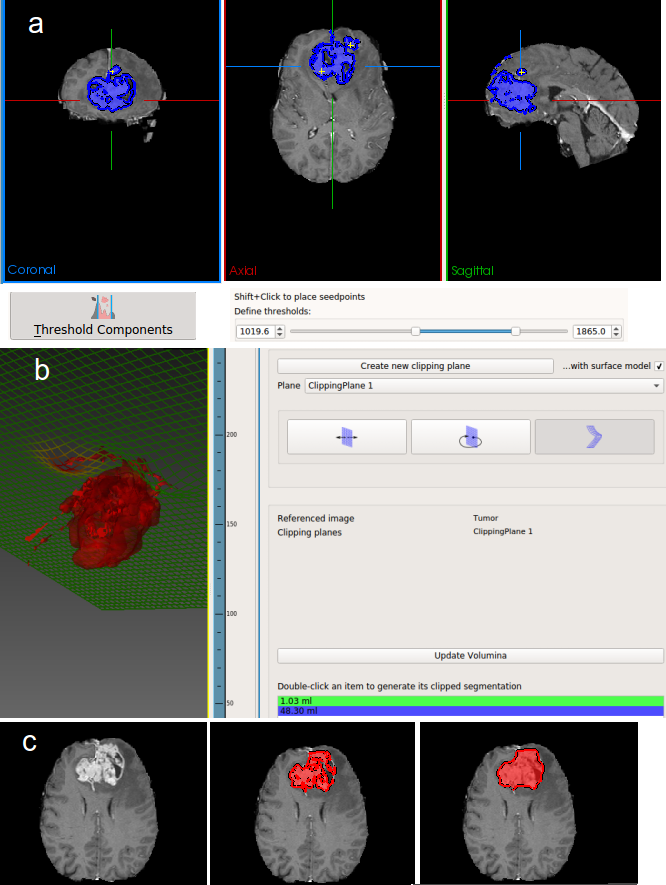}
	\caption{\textbf{Demonstration of a user-friendly semiautomatic segmentation protocol.} a) The Threshold Components tool allows the user to set an intensity threshold and multiple seedpoints (yellow crosses). Regions within the threshold and connected to the seedpoints are selected. b) The deformable clipping plane allows 3D correction of the segmentation, typically useful for removing leaked regions. c) Original image, clipped image, and final morphological operations such as Closing and Fill Holes smooth the enhancing segmentation into a Core segmentation.}
	\label{fig:protocol}
\end{figure}

	\subsubsection{Thresholding}
	A new segmentation label was created for the T1c image, and the Segmentation plugin view is opened. Selecting the newly implemented Threshold Components tool, seed point(s) were placed in enhancing region(s), and the lower threshold adjusted until the apparent hyper-intensities were all included (Fig. 1a). If the tumor was grossly non-enhancing, another label was created where seed point(s) were placed in hypo-intense regions, and the upper threshold adjusted.

	\subsubsection{Cropping}
	Regions of normal brain tissue that were erroneously included as part of the Thresholding process were removed in two ways. Precise exclusion of regions was done via the Clipping Plane tool, where up to 6 deformable 3D surfaces were superimposed on the segmentation. This allowed the generalized separation of erroneous regions from the tumor body (Fig. 1b), a new feature added in MITKats. For gross corrections, the existing Image Cropping tool could be used instead. A 3D rectangular bounding box was graphically defined and used to mask and overwrite the original segmentation. Finally, if the true tumor contained only one connected component, either of these methods could be followed up with the Picking tool to exclude erroneous regions that were severed from the tumor.
	 
	\subsubsection{Smoothing and Filling}
	The segmentation was smoothed via the Closing tool, a part of the existing set of Morphological Operations, typically with a radius of 2. This label was saved as the Active component. For the Core volume, the non-enhancing label could be joined to this component via the Union tool, if applicable. To include areas of necrosis, the Closing and Fill Holes tools were used (both typically with radius of 10), and this new label was saved as the Core component (Fig. 1c).

	\subsubsection{Whole Tumor}
	For segmenting the Whole tumor component of the BRATS dataset, the above protocol for obtaining the Core region was repeated for the FLAIR image.

\subsection{Accuracy and Speed Analysis}
 The accuracy of MITKats segmentations was assessed by comparison to the reference segmentations provided by BRATS and St. Olav's datasets. In BRATS, the references were ground truths fused from 4 expert manual annotations. Each of the Whole, Core, and Active components were compared to their reference counterparts via Dice score \citep{Dice1945} as well as calculated tumor volume. The mean inter-rater Dice scores from BRATS were used as controls. This was because the fused ground truths were derived from the individual expert annotations, artificially bolstering the Dice score between any given expert and the fused standard. 
 
 For comparing data in the St. Olav's cohort, the MITKats segmentations were assessed via Dice score and volume to all 12 reference segmentations (3 softwares $ \times $ 2 raters $ \times $ 2 repetitions). As a control, Dice scores were calculated only for pairs of reference segmentations created by different raters. The time to create the Core component of the MITKats segmentation was compared against the times reported in the dataset.

\section{Results}
Aggregate Dice scores and volume estimates of MITKats segmentations with respect to BRATS ground truths for each tumor region are shown in Table \ref{tab:brats}. The Dice similarity of MITKats segmentations compared to the ground truth was equivalent to inter-rater variability for the Whole and Active tumor regions, but was worse for the Core region. Volume measurements averaged over all regions were 94\% of those estimated by the ground truth, with a mean fractional error of 18\%.

\begin{table}[h]
	\begin{center}
		\begin{tabular}{|c|c|c|c|c|}
			\hline
			\multicolumn{2}{|c|}{Tumor Compartment}  & Whole & Core & Active \\
			\hline
			\multirow{2}{*}{Dice Score (\%)} & MITKats v. BRATS & 88$\pm $5 & 84$\pm $10 & 77$\pm $14 \\
			\cline{2-5}
			 & BRATS Inter-rater & 88$\pm $2 & 93$\pm $3 & 74$\pm $13 \\
			\hline
			\multicolumn{2}{|c|}{Volume Ratio} &  0.96 & 1.06 & 0.81 \\
			\hline
			\multicolumn{2}{|c|}{Volume Relative Error} &  10\% & 25\% & 20\% \\
			\hline
			
		\end{tabular}
	\end{center}
\caption{\textbf{Segmentation accuracy approaches inter-rater agreement in the Brain Tumor Segmentation Challenge.} Twenty high-grade glioma cases were segmented using MITKats and compared against ground truths from BRATS. Three regions (Whole, Core, and Active) were segmented for each patient, and the mean Dice scores ($\pm$ standard deviation) are shown. The volumes of each region were also compared to the ground truth.}
\label{tab:brats}
\end{table}

\begin{figure}[htp!]
	\includegraphics[width=\textwidth]{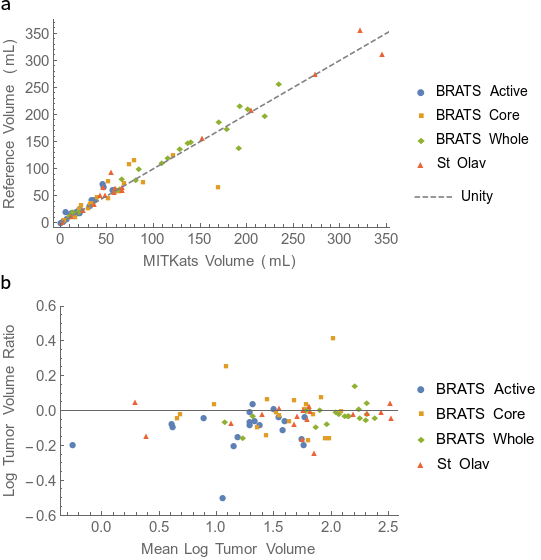}
	\caption{\textbf{Comparison of segmented volumes by tumor region.} a) Volumes calculated using MITKats are compared against their reference standards for each tumor component. (St. Olav segmentations designated only a single Core-like component.) b) A logarithmic Bland-Altman plot  is shown for different regions of segmentations, comparing the ratio of segmented volumes to the averages of the base 10 logarithms of tumor volume (in mL). There is a tendency for MITKats to underestimate tumor volume at low sizes, particularly in the Active tumor component.}
	\label{fig:volume}
\end{figure}

A comparison of estimated volumes for each case across both datasets is shown in Figure \ref{fig:volume}, including a logarithmic Bland-Altman analysis. While the volumes segmented by MITKats were similar to reference segmentations, they tended to be underestimated, particularly for the Active tumor region of the BRATS dataset.

\begin{figure}[htp!]
	\includegraphics[width=\textwidth]{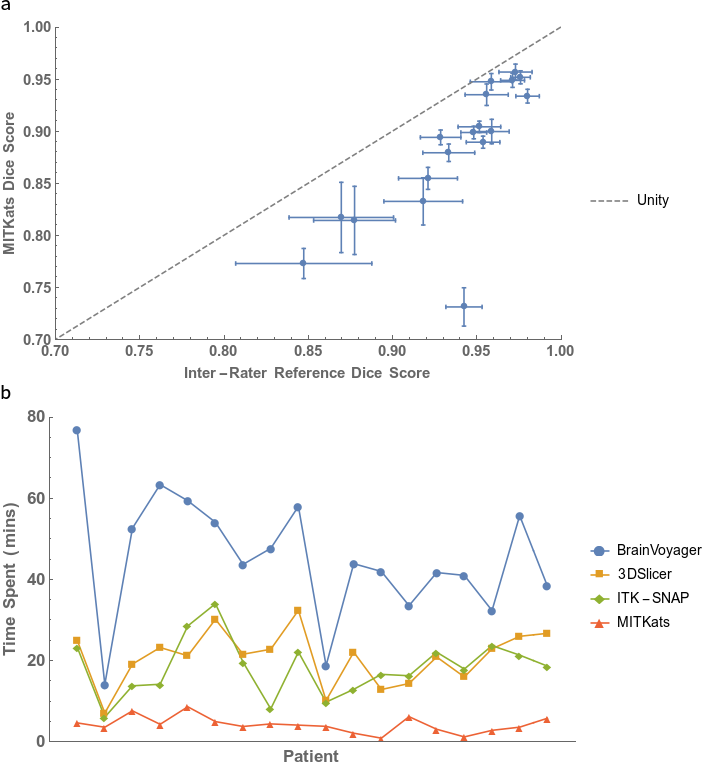}
	\caption{\textbf{MITKats is faster than other segmentation softwares, while approaching optimal accuracy.} a) Eighteen glioblastoma cases were segmented using MITKats and compared against the segmentations performed in \citet{Fyllingen2016}. Points represent the average Dice score when compared to all other segmentations performed by different raters. Error bars represent standard deviation.  b) The time required for segmenting via MITKats is compared to those using BrainVoyager, 3DSlicer, and ITK-SNAP (mean of 2 trials each). 
		}
	\label{fig:speed}
\end{figure}

The Core component of the MITKats segmentation was pairwise compared to segmentations performed by other softwares in the St. Olav dataset. The average Dice scores of MITKats segmentations compared to each of the reference segmentations was 0.88, compared against their inter-rater agreement of 0.94 (Figure \ref{fig:speed}a).

The time for segmentation is also compared to the St. Olav's dataset in Figure \ref{fig:speed}b. The speed of MITKats as compared to those reported was an average of 4, 5, and 11 times faster than ITK-SNAP, 3D Slicer, and BrainVoyager, respectively. The typical Core segmentation using MITKats was 4.2 $\pm $ 2.0 minutes on the St. Olav's dataset and 4.0 $\pm $ 3.1 minutes on the BRATS dataset, where uncertainties represent standard deviation.

\section{Discussion and Conclusion}
To our knowledge, this work is the first to validate the use of the MITK environment for the segmentation of brain tumors. Our modification MITKats provides fast and accurate segmentation, combining a semiautomatic tool with flexible 3D editing, all wrapped in a user-friendly GUI. Benchmarking its accuracy against BRATS, it achieved a performance equal to inter-rater variability across Whole and Active tumor regions. The Core tumor region was somewhat lacking, perhaps due to our protocol not using T1 images, thus having different definitions for the component. Benchmarking its speed against the St. Olav's dataset, MITKats was over 4 times faster than its quickest competitor. While its accuracy was somewhat lower than inter-rater variability, the reference segmentations were also semi-automatically derived, and therefore not necessarily the ground truth. Finally, MITK is an open source toolkit which encourages the free use and continued development of this software.

The scope of this work has some limitations with the protocol and datasets used. This pipeline works best for T1c and FLAIR imagery, because T1 without contrast has limited enhancement, and T2 enhancements are often the same intensity as cerebrospinal fluid. We hope to provide implementation for other modalities in the future as well as simultaneous multimodal analysis, which may benefit the accuracy of segmentations \citep{Dupont2016}. Providing support for perfusion and diffusion weighted images may also be particularly important for response assessment \citep{Huang2015}. 

One note is that the BRATS dataset contained not just glioblastomas but also anaplastic astrocytomas, which are intermediates between GBMs and lower grade gliomas (LGGs). Given the satisfactory accuracy of segmenting this mixed dataset, this provides encouragement that MITKats could be extended to LGGs, which are typically harder to segment \citep{Akkus2015}. Eventual implementation and validation for LGGs and tumors outside the brain would be useful. Another point about the BRATS dataset is that the images were already skull-stripped, making the segmentation process somewhat easier. Integrating skull-stripping into our pipeline would be helpful for the user.

It is our belief that speed is not sufficiently emphasized in the current benchmarking of segmentation algorithms. While many studies report average segmentation times, the variability among different cases can be large. Within the datasets examined in this work, the fastest segmentation took less than a minute, while the slowest took almost 11 minutes. Therefore, the individual listing of segmentation times is helpful for the direct comparison of algorithm speeds. We encourage future investigators to record both raw segmentations and operator time for each case to further advance this field.

%\section*{References}

\bibliography{mybibfile}

\section*{Acknowledgments}
We would like to thank Anthea Monod and the rest of the Rabadan lab for their helpful comments and feedback. A.X.C. is funded by NIH grant T32GM007367 as part of the Medical Scientist Training Program (MSTP). R.R. acknowledges funding from the NIH (U54 CA193313, R01 CA185486, R01 CA179044).

%\section*{Vitae}
%Andrew Chen (A.X.C.) is an MD/PhD candidate at Columbia University. Previously, he was an undergraduate at MIT, where he obtained degrees in Physics and Biology. He is interested in applying computational methods to medical image processing and cancer evolution.
%
%Ra\'{u}l Rabad\'{a}n (R.R.) is an Associate Professor in the Department of Systems Biology and Biomedical Informatics at Columbia University. He is the director of the Center for Topology of Cancer Evolution and Heterogeneity. Previously, he was the Martin A. and Helen Chooljian Member at The Simons Center for Systems Biology at the Institute for Advanced Study in Princeton, New Jersey. His current interest focuses on uncovering patterns of evolution in biological systems—in particular, RNA viruses and cancer.

\end{document}